\documentclass[groupedaddress,showpacs]{revtex4}
\usepackage[dvips]{graphicx}
\begin{document}
\title{Spin quantum computation in silicon nanostructures}
\author{S. Das Sarma}
\affiliation{Condensed Matter Theory Center, Department of Physics, University
of Maryland, College Park, MD 20742-4111}
\author{Rogerio de Sousa}
\affiliation{Department of Chemistry and Pitzer Center for Theoretical 
Chemistry, University of California, Berkeley, CA 94720-1460}
\author{Xuedong Hu}
\affiliation{Department of Physics, University at Buffalo, SUNY, Buffalo, NY
14260-1500}
\author{Belita Koiller}
\affiliation{Instituto de F\'{\i}sica, Universidade Federal do Rio de
Janeiro, Cx. Postal 68.528, Rio de Janeiro, 21945-970, Brazil}

\begin{abstract}
Proposed silicon-based quantum-computer architectures have attracted
attention because of their promise for scalability and their potential
for synergetically utilizing the available resources associated with the
existing Si technology infrastructure.  Electronic and nuclear spins of 
shallow donors (e.g. phosphorus) in Si are ideal candidates for 
qubits in such proposals because of their long spin coherence times due to
their limited interactions with their environments.  For these spin qubits,
shallow donor exchange gates are frequently invoked to perform two-qubit 
operations.  We discuss in this review a particularly important spin
decoherence channel, and bandstructure effects on the exchange gate
control.  Specifically, we review our work on donor electron spin spectral
diffusion due to background nuclear spin flip-flops, and how isotopic
purification of silicon can significantly enhance the electron spin dephasing
time.  We then review our calculation of donor electron exchange coupling in
the presence of degenerate silicon conduction band valleys.  We show that
valley interference leads to orders of magnitude variations in electron
exchange coupling when donor configurations are changed on an atomic scale. 
These studies illustrate the substantial potential that donor
electron/nuclear spins in silicon have as candidates for qubits and
simultaneously the considerable challenges they pose.  In particular, our
work on spin decoherence through spectral diffusion points to the possible
importance of isotopic purification in the fabrication of scalable solid
state quantum computer architectures.  We also provide a critical comparison
between the two main proposed spin-based solid state quantum computer
architectures, namely, shallow donor bound states in Si and localized quantum
dot states in GaAs.
\end{abstract}
\maketitle

\pagebreak


\section{Introduction}

During the past decade, the study of quantum computing and quantum information
processing has generated widespread interest among physicists from areas
ranging from atomic physics, optics, to various branches of condensed matter
physics \cite{Reviews,nielsen00}.  The key thrust behind the rush toward a
working quantum computer (QC) is the development of a quantum algorithm that
can factorize large numbers exponentially faster than any available classical
algorithm \cite{Shor}.  This exponential speedup is due to the intrinsic
quantum parallelism in the superposition principle and the unitary evolution
of quantum mechanics.  It implies that a computer made up of entirely quantum
mechanical parts (qubits), whose evolution is governed by quantum mechanics,
would be able to carry out prime factorization of large numbers that is
prohibitively time-consuming in
classical computation, thus revolutionizing cryptography and information
theory.  Since the invention of Shor's factoring algorithm, it has also been
shown that error correction can be done to a quantum system \cite{error}, so
that a practical QC does not have to be forever perfect to be useful, as long
as quantum error corrections can be carried out on the QC.  These two key
mathematical developments have led to the creation of the new
interdisciplinary field of quantum computation and quantum information.

Many physical systems have been proposed as candidates for qubits in a QC. 
Among the more prominent examples are electron or nuclear spins in
semiconductors \cite{igorrmp,SSReviews}, including electron spin in
semiconductor quantum dots \cite{Exch,Imam} and donor electron or nuclear
spins in semiconductors \cite{Kane,Privman,Vrijen}.  The donor-based QC
schemes are particularly interesting because all donor electron wavefunctions
in a semiconductor are identical, and because doping makes a natural
connection
between quantum mechanical devices and the more traditional microelectronic
devices: Doping in semiconductors has had significant technological impact
for the past fifty years and is the basis of the existing microelectronics
technology.  As transistors and integrated circuits decrease in size, the
physical properties of the devices are becoming sensitive to the actual
configuration of impurities \cite{voyles}.  In this context, the important
proposal of donor-based silicon quantum computer (QC) by Kane \cite{Kane}, in
which the nuclear spins of the monovalent $^{31}$P impurities in Si are the
qubits, has naturally created considerable interest in revisiting all
aspects of the donor impurity problem in silicon, particularly in the
Si:$^{31}$P system.

In principle, both electron spin and orbital degrees of freedom can be used as
qubits in semiconductor nanostructures.  For example, electron orbital
dynamics is quantized into discrete ``atomic-like'' levels (on meV energy
scale with a Bohr radius of the order of 10 nm) in semiconductor quantum
dots, and two such quantized quantum dot levels could form the quantum
two-level system needed for a qubit \cite{Tanamoto}.  A great advantage of
such orbital (or equivalently, charge) qubits is that qubit-specific
measurements are relatively simple since one is essentially measuring single
charge states, which is a well-developed experimental technique through
single-electron transistors (SET) or equivalent devices \cite{SCT}.  A major
disadvantage of solid state charge qubits is that these orbital states are
highly susceptible to interactions with the environment 
(which, in particular, contains all the stray or unintended
charges inevitably present in the device), and the decoherence time is
generally far too short (typically picoseconds to nanoseconds) for quantum
error correction to be useful.  A related problem is that inter-qubit
coupling, which is necessary for the implementation of two-qubit gate
operations essential for quantum computation, is often the long-range dipolar
coupling for charge qubits.  This makes it difficult to scale up the
architecture, since decoherence grows with the scaling-up as more and more
qubits couple to each other via the long-range dipolar coupling.  Additional
decoupling techniques have to be applied to ensure selective gate operations
\cite{Dykman}, and to control decoherence due to long range couplings
\cite{DD_dipolar}.  Since scalability is thought to be the main advantage of
solid state QC architectures, little serious attention has so far been paid
to orbital qubit based quantum computation in semiconductor nanostructures
due to its probably unscalable decoherence and entanglement properties.  

Spin qubits in semiconductor nanostructures have complementary advantages (and
disadvantages) compared with charge qubits based on quantized orbital states. 
A real disadvantage of spin qubits is that a single electron spin is
difficult to measure, although there is no fundamental principle against
the measurement of a Bohr magneton.  The great advantage of spin qubits
is the very long spin coherence times, which even for electron spins can be
milliseconds (microseconds) in silicon (GaAs) at low temperatures.  This six
orders of magnitude coherence advantage in spin qubits over charge qubits has
led to electron spin qubits in GaAs quantum dots and in P donor levels in
silicon (as well as SiGe quantum structures) as the QC architectures of
choice for the solid state community.  In addition to the coherence
advantage, spin qubits also have a considerable advantage that the exchange
gate, which provides the inter-qubit coupling, is exponentially
short-ranged and nearest-neighbor in nature, thus allowing precise control and
manipulation of two-qubit gates.  There is no fundamental problem arising from
the scaling-up of the QC architecture since exchange interaction couples only
two nearest-neighbor spin qubits independent of the number of qubits.  In
this review we provide a brief perspective on spin qubits in silicon with
electron spins in shallow P donor levels in Si being used as qubits.

Although experimental progress in semiconductor-based solid state QC schemes
has been slow to come during the past five years, they are still often
considered promising in the long term because of their perceived scalability
advantages.  After all, the present computer technology is based on
semiconductor integrated circuits with ever smaller feature size.
Therefore, semiconductor nanostructure-based QC architectures should in
principle be scalable using the existing microelectronics technology. 
However, it still remains to be demonstrated whether and how the available
(classical) semiconductor technology can help the scaling up  of a quantum
coherent QC
architecture.  For the spin qubits in silicon, for example, the key issues
include clarifications of spin quantum coherence properties in the solid
state environment, physical approaches to manipulate and entangle spins,
fabrication of devices with atomic-scale precision, and measurement of
single spins.  In the following, we review our work on two of these important
issues: donor electron spin coherence and spin interaction in silicon.

\section{Spin coherence in silicon nanostructures}

Before the seminal concept of quantum error correction was introduced in 1995
\cite{error}, it was widely believed that quantum computation, even as a
matter of
principle, is quite impossible since all quantum states decohere due to
interaction with the environment, and such decoherence was thought to be
fatal to QC operations.  Although the quantum error correction principle has
shown that a certain degree of decoherence can be corrected in QC algorithms,
one
still has severe limits on the amount of tolerable decoherence for feasible
QC operations.  The question of decoherence is therefore of paramount
importance for any form of quantum computing architecture, including the
spin-based proposals.  Estimates of exchange coupling and various analysis
of adiabatic operation of the exchange gates in quantum dots \cite{HD_gate}
suggest that the exchange gate duration is unlikely to be shorter than
$100$~ps--adiabaticity is likely to be compromised for shorter pulse times
leading to loss of qubit fidelity.  Hence spin coherence times at least of
the order of microseconds are necessary to satisfy the criteria of
fault-tolerance ($10^4$ reliable quantum gate operations during the coherence
time \cite{nielsen00}).  Non-ideal situations encountered in real devices
will probably require even longer coherence times.  Here we argue that the
possibility of using nuclear-spin-free samples of silicon through isotopic
purification together with the small spin-orbit coupling in silicon makes Si
one of the most promising host materials with respect to electron spin
coherence.  As we show below, both
adiabaticity (i.e. ``slow'' gate operation) and fault tolerance (i.e.
``fast'' gate operation) are more easily satisfied in Si donor electron spins
than in GaAs quantum dots.

The simplest description of electron spin coherence is to consider
characteristic time scales $T_1$ and $T_2$ \cite{abragam}, which are
respectively defined as the decay times of spin magnetization parallel (the
so-called longitudinal spin relaxation time $T_1$) and perpendicular (the
so-called transverse spin dephasing time $T_2$) to an external magnetic
field.  All processes contributing to $T_1$ require energy exchange with the
lattice (via phonon emission, for example).  These $T_1$ relaxation processes
also contribute to the dephasing time $T_2$, hence the inequality $T_2\leq 2
T_1$ \cite{yafet63}.  However, often one finds $T_2\ll T_1$ due to the
importance
of dephasing mechanisms that do not involve energy relaxation, thus
contributing exclusively to $T_2$ decay.  This is exactly the case for a Si:P
donor electron spin.  In this system $T_1$ can be as long as $10^3$~seconds
\cite{feher59b} (measured for $B=0.3$~T -- we note that $T_1\propto B^{-5}$
at low temperatures due to phonon matrix element and phase space
considerations
\cite{feher59b}), while $T_2$ detected by spin echo decay is of the order of
$10^{-4}$~seconds \cite{chiba72}.  The reason why $T_1$ is so long in silicon
nanostructures is two fold: (1) Spin flips mediated by spin-orbit coupling
and the electron-phonon interaction are strongly suppressed for localized
electrons \cite{pines57} due to energy-momentum conservation constraints, and
(2) spin-orbit coupling in silicon is significantly weaker than in other
popular semiconductors \cite{tahan02} (such as III-V compounds, where
$T_1$ is of the order of nanoseconds for conduction electrons
\cite{igorrmp}).  This observation, together with the fact that one should
expect $T_2=2 T_1$ in isotopically purified $^{28}$Si (see below), indicates
that silicon is one of the most attractive materials for coherent spin
manipulation.

It was discovered a long time ago \cite{herzog56} that the dipolar fluctuation
of lattice nuclear spins is responsible for electron spin phase fluctuations, 
leading to significant suppression of $T_2$, an effect usually denoted
spectral diffusion.  Here each nuclear spin produces a different hyperfine
shift in the electron Zeeman frequency.  Because the nuclei are coupled to
each other via magnetic dipolar interaction leading to nuclear spin flip-flop
transitions, the total hyperfine field at the localized electron fluctuates
and leads to time-dependent noise in the electron spin Zeeman frequency and
its consequent decoherence.  Below we summarize a microscopic theory for this
spectral diffusion effect which was proposed by two of us
\cite{desousa03a,desousa03b} and recently verified experimentally
\cite{abe04,tyryshkin03}. 

\begin{figure}
\begin{center}
\includegraphics[width=3.4in]{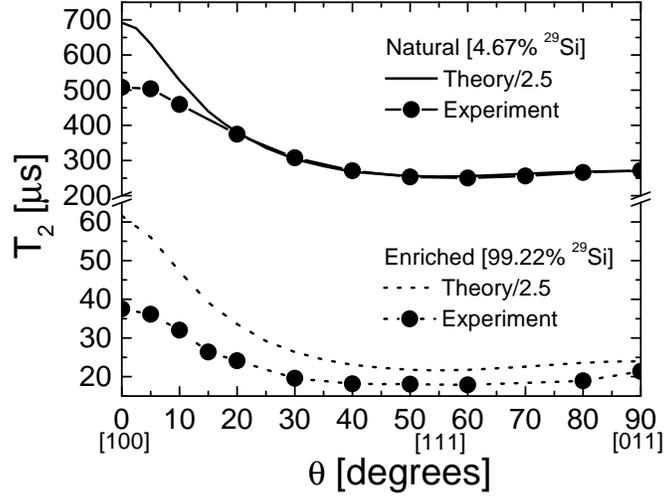}
\caption[]{Electron spin coherence time for a Si:P donor. $\theta$ denotes the
angle between the external magnetic field and the [100] crystallographic
direction. The agreement between theory and experiment is within a factor of
3 (experimental data from Ref.~\cite{abe04}).}
\label{fig_exp}
\end{center}
\end{figure}

Silicon has three stable isotopes: $^{29}$Si has nuclear spin quantum number
$I=1/2$ and a natural abundance equal to 4.68\%.  $^{28}$Si and $^{30}$Si are
spin-$0$ and hence do not contribute to spectral diffusion. Nuclear isotope
engineering \cite{abe04,tyryshkin03} can systematically reduce the amount of
$^{29}$Si (``the isotopic purification''), thus raising an important
theoretical question on how $T_2$ depends on the fraction of $^{29}$Si
present in the sample.  (We note that $T_2$ due to spectral diffusion becomes
infinitely long as the fraction of $^{29}$Si in the sample approaches zero.) 
To address this question, we consider a truncated Hamiltonian for the
electron-nuclear spin evolution,
\begin{eqnarray}
{\cal H} & = & \gamma_{S}B S_{z}-\gamma_{I}B\sum_{n}I_{nz}\nonumber\\
&& + \sum_{n}A_{n}I_{nz}S_{z} - 4\sum_{n<m}b_{nm}I_{nz}I_{mz}\nonumber\\
&& + \sum_{n<m}b_{nm}(I_{n+}I_{m-}+I_{n-}I_{m+}).
\label{htot}
\end{eqnarray}
Here $S_z$ is the z-component of the electron spin operator, and $\gamma_S$
is the electron gyromagnetic ratio. ${\bf
I}_n$ is the nuclear spin operator for a $^{29}$Si isotope located at
position ${\bf R_n}$, $A_n$ is the hyperfine shift produced by this nucleus,
\begin{equation}
A_{n}= \gamma_{S}\gamma_{I}\hbar \frac{8\pi}{3}|\Psi({\mathbf R_{n}})|^{2},
\end{equation}
and $b_{nm}$ is the dipolar coupling between two such nuclei, 
\begin{equation}
b_{nm}=-\frac{1}{4}\gamma_{I}^{2}\hbar\frac{1-3\cos^{2}\theta_{nm}}{R_{nm}^{3}},
\label{bnm}
\end{equation}
which has an important dependence with respect to the angle $\theta_{nm}$
between the magnetic field and the vector linking the two nuclei
${\bf R}_{n}-{\bf R}_{m}$. In Eq.~(\ref{htot}) we neglect the off-diagonal
hyperfine coupling ($S_{+}I_{n-}+\rm{h.c.}$), which can be shown to play no
role at magnetic fields higher than the Overhauser field ($B>\sum_n
A_n\sim 10-1000$~G in silicon \cite{desousa03b}).  The last term of
Eq.~(\ref{htot}) is responsible for flip-flop processes between two nuclear
spins, which can be studied using an inverse temperature expansion. The rate
for a flip-flop process is then given by \cite{desousa03b}
\begin{equation}
\Gamma_{nm} =\frac{2\pi b_{nm}^2}{\sqrt{2\pi \kappa_{nm}^2}}
\exp{\left(-\frac{\Delta_{nm}^2}{2\kappa_{nm}^2}\right)}.
\label{tnm}
\end{equation}
Here $\Delta_{nm}=|A_n-A_m|/2$ is the frequency shift felt by the
electron when one flip-flop event takes place, while the linewidth for
flip-flop is given by 
\begin{equation}
\kappa_{nm}^{2}=4\sum_{i\neq n,m}
\left(b_{ni}-b_{mi}\right)^2 \,.
\label{kappa}
\end{equation}
Nuclear spin flip-flop can only take place if the nuclear system can rearrange
itself to compensate for the energy cost $\Delta_{nm}$.  Hence pairs of
nuclei satisfying $\Delta_{nm} \gg \kappa_{nm}$ are inert.  Since these are
usually located close to the center of the donor (where $A_n$ is larger),
they form a frozen core which does not contribute appreciably to spectral
diffusion. 

\begin{figure}
\begin{center}
\includegraphics[width=3.4in]{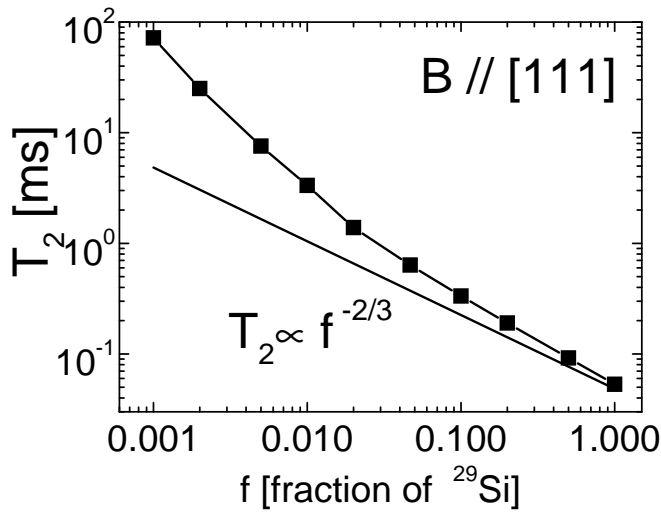}
\caption[]{Dependence of Si:P electron spin coherence time (as measured by
Hahn echo decay) with the fraction of $^{29}$Si nuclear spins. This figure
reveals isotopic purification as an effective tool for material optimization
of $T_2$.}
\label{fig_si29}
\end{center}
\end{figure}

An approximate theory for spectral diffusion decay can be obtained by assuming
that the most important processes contributing to nuclear dipolar fluctuation
are uncorrelated flip-flop transitions with Markovian dynamics.  The
$\pi/2-\tau-\pi-\tau-echo$ envelope arising due to the stochastic fluctuation
of one pair of nuclei $n,m$ is then determined to be \cite{desousa03b}
\begin{eqnarray}
v_{nm}(2\tau)&=&R_{nm}^{-2} \textrm{e}^{-2\tau\Gamma_{nm}} \biggl\{
\Gamma_{nm}^{2}\cosh{(2\tau R_{nm})} \nonumber \\ 
&&+\Gamma_{nm} R_{nm} \sinh{(2\tau R_{nm})}-\Delta_{nm}^{2}\biggr\},
\label{vnm}
\end{eqnarray}
where $R_{nm}^{2}=T_{nm}^{-2}-\Delta_{nm}^{2}$.  We then consider the
contribution of all nuclear spins through an averaged product
\begin{equation}
v(2\tau)=\prod_{n<m}\left[(1-f^2)+f^{2} v_{nm}(2\tau)\right],
\label{vtot}
\end{equation}
where $f$ is the fraction of $^{29}$Si nuclear spins present in the lattice.
Equation~(\ref{vtot}) allows realistic calculations of echo decay due to
spectral diffusion without using any fitting parameters.  All we need is an
appropriate wave function for the electron (we use the Kohn-Luttinger state
for a Si:P impurity) which in turn determines the hyperfine shifts
$\Delta_{nm}$
and the corresponding fluctuation rates $\Gamma_{nm}$ (other constants such as
$\gamma_S$, $\gamma_I$ are readily available from the NMR literature).  For
natural silicon, the calculated $T_{2}$ is $2.5$ times larger than the
measured value \cite{chiba72}, a reasonable agreement
in the spin relaxation literature \cite{igorrmp}.  In addition we
were able to predict an important orientation dependence for $T_2$ (see
Fig.~\ref{fig_exp}).  As the direction of the external magnetic field is
changed with respect to the silicon diamond structure $T_2$ increases by as
much as a factor of three.  Recently, this prediction was verified
experimentally for natural silicon \cite{tyryshkin03,abe04} and for a
$^{29}$Si enriched sample \cite{abe04} (Fig.~\ref{fig_exp}).  The calculated
dependence of $T_2$ on the fraction of $^{29}$Si is shown in
Fig.~\ref{fig_si29}, establishing isotopic purification as an efficient way
to achieve the longest possible electron spin coherence times for QC use.
Recently A. M. Tyryshkin and collaborators \cite{tyryshkin03} performed echo
spectroscopy in a sample with less than 0.005\% $^{29}$Si isotopes.  The
residual $T_2>60$~ms was attributed to the lack of coherence in the applied
ESR pulses, and not to the spectral diffusion process.

\begin{figure}
\includegraphics[width=3.in]{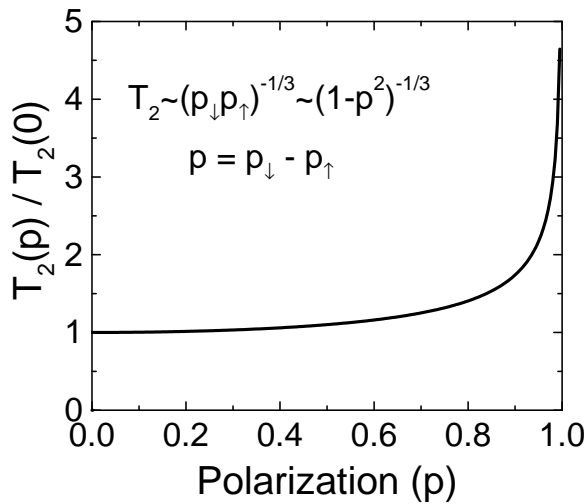}
\caption{Depicts the dependence of coherence time $T_2$ with the
  nuclear spin polarization $p$. Very high nuclear polarization
  ($p>0.99$) is required to enhance $T_2$, a condition which is very
  hard to achieve experimentally.
\label{t2pol}}
\end{figure}

Isotopic purification can also be used in germanium, where the active nuclear
spin is $^{73}$Ge ($I=9/2$, 7.73\% natural abundance). On the other hand the
important class of III-V semiconductors (GaAs, GaSb, InAs, InSb) has no known
$I=0$ nuclear isotopes.  Because echo decay has never been measured for
localized spins in these materials, the role of theory becomes extremely
important.  Recently we predicted $T_2\sim 1-100$~$\mu$s
\cite{desousa03b,desousa03a} for localized electron spins in III-V
semiconductor quantum dots, where the development of single spin detection
techniques promises a bright future for coherent spin
manipulations \cite{elzerman04}.  There are currently two
proposals to reduce the effect of strong spectral diffusion associated with
the absence of $I=0$ isotopes in III-V materials.  The first relies on
substantial nuclear polarization \cite{burkard99} (very high nuclear
polarization is required to suppress the flip-flop processes that lead to
spectral diffusion \cite{desousa03a}.  The qualitative dependence of $T_2$ on
nuclear polarization can be derived from the short $\tau$ limit of
Eqs.~(\ref{vnm}) and (\ref{vtot}). In this case we have \cite{desousa03a}
\begin{equation}
\left(\frac{1}{T_2}\right)^3\propto p_{\uparrow}p_{\downarrow}\propto
(1-p^2),
\end{equation}
where the nuclear spin polarization is given by
$p=p_{\uparrow}-p_{\downarrow}$.  This dependence implies $T_2$ is
enhanced significantly only if $p$ is very close to 1
(Fig.~\ref{t2pol}).  For example, a nuclear polarization of
$p=99.5$\% increases $T_2$ only by a factor of 5).  The other 
approach to reduce spectral diffusion requires
successive applications of $\pi$ pulses (Carr-Purcell-Meiboom-Gill sequence)
at a rate proportional to the square of the nuclear spin quantum number
\cite{desousa04}.  Both these proposals require considerable overhead, and
make evident the advantages of silicon or germanium from the perspective of
coherent spin manipulation.

We make two brief final comments before concluding our discussion of spin
decoherence.  The first comment is that we believe the spectral diffusion
mechanism to be the ultimate decoherence mechanism for localized electron
spins in semiconductor nanostructures since it cannot be eliminated in a
given sample by reducing the sample temperature as the nuclear spin energy
scale is extremely small (sub-millikelvins).  Thus, understanding spectral
diffusion effects in spin-based solid state QC architectures is one of the
most important physics issues in the context of electron spin dynamics in
quantum computers.  From this narrow perspective, silicon QC has significant
advantages over GaAs QC due to the unique role of isotopic purification in
silicon.  The second comment is that our statement about the spin coherence
time in GaAs nanostructures (or more generally, in III-V materials) being
`short' ($T_2 \sim 1-100$~$\mu$s) should be understood in the proper context
of a spectral diffusion comparison with the corresponding silicon coherence
where $T_2 \sim 1-100$ ms.  In this context, recent claims in the literature
of very long spin
coherence times ($\sim$ 1-100 ns) in GaAs for conduction electrons are in
fact extraordinarily short for the purpose of spin quantum computation.

\section{Donor Electron Exchange in Silicon}

An important issue in the study of donor-based Si QC architecture is coherent
manipulations of spin states as required for the quantum gate operations.  In
particular, two-qubit operations, which are required for a universal QC,
involve precise control over electron-electron exchange
\cite{Exch,Kane,Vrijen,HD} and electron-nucleus hyperfine interactions (for
nuclear spin qubits).  Such control can presumably be achieved by fabrication
of donor arrays with accurate positioning and surface gates whose potential
can be precisely controlled \cite{Obrien,encapsulation,implant,schenkel03}.  
However, electron exchange in bulk silicon has spatial
oscillations \cite{Andres,KHD1} on the atomic scale due to valley 
interference arising from the particular six-fold degeneracy of the bulk Si
conduction band.  These exchange oscillations place heavy burdens on device
fabrication and coherent control \cite{KHD1}, because of the very high
accuracy and tolerance requirements for placing each donor inside the Si unit
cell, and/or for controlling the external gate voltages.  

The potentially severe consequences of the exchange-oscillation problem for
exchange-based Si QC architecture motivated us and other researchers to
perform theoretical studies going beyond some of the simplifying
approximations in the formalism adopted in Ref.~\onlinecite{KHD1}, and
incorporating perturbation effects due to applied strain\cite{KHD2} or gate
fields \cite{wellard03}.  These studies, all performed within the standard
Heitler-London (HL) formalism \cite{slater}, essentially reconfirm the
originally reported difficulties regarding the sensitivity of the electron
exchange coupling to precise atomic-level donor positioning, indicating that
they may not be completely overcome by applying strain or electric fields. 
The sensitivity of the calculated exchange coupling to donor relative 
position originates from interference between the plane-wave parts of the six
degenerate Bloch states associated with the Si conduction-band minima. More
recently \cite{KCHD} we have assessed the robustness of the HL approximation
for the two-electron donor-pair states by relaxing the phase pinning at donor
sites, which could in principle eliminate the oscillatory exchange behavior.
Within this more general theoretical scheme, the {\em floating-phase} HL
approach, our main conclusion is that, for all practical purposes, the
previously adopted HL wavefunctions are robust, and the oscillatory behavior
obtained in Refs.~\onlinecite{KHD1,KHD2,wellard03} persists (quantitatively)
in the more sophisticated theory of Ref.~\cite{KCHD}.

In what follows, we first review the main results leading to the exchange
oscillation behavior qualitatively described above.  We then consider two
substitutional donors in bulk Si, and present a systematic statistical study
of the correlation between the relative position distributions and the
resulting exchange distributions.  We also show that strain may partially
alleviate the exchange oscillatory behavior, but it cannot entirely overcome
it.

\subsection{Donor Electron Exchange in Relaxed Bulk Silicon}

We describe the single donor electron ground state using the effective mass
theory.  The bound donor electron Hamiltonian for an impurity at site ${\bf
R}_0$ 
is written as
\begin{equation}
{\cal H}_0={\cal H}_{SV}+{\cal H}_{VO} \,.
\label{eq:h0}
\end{equation}
The first term, ${\cal H}_{SV}$, is the single-valley Kohn-Luttinger
Hamiltonian \cite{Kohn}, 
which includes the single particle kinetic energy, the Si periodic potential,
and the screened Coulomb perturbation potential produced by the impurity ion
\begin{equation}
V({\bf r})=-\frac{e^2}{\epsilon|{\bf r}-{\bf R}_0|} \,. 
\label{eq:coul}
\end{equation}
For shallow donors in Si, we use the static dielectric constant $\epsilon =
12.1$.  The second term of Eq.~(\ref{eq:h0}), ${\cal H}_{VO}$, represents the 
inter-valley scattering effects due to the presence of the impurity. 

The donor electron eigenfunctions are written in the basis of the six
unperturbed Si band edge Bloch states $\phi_\mu = u_\mu(\bf r) e^{i {\bf
k}_{\mu}\cdot {\bf r}}$ [recall that the conduction band of bulk Si has six
degenerate minima $(\mu=1,\ldots,6)$, located along the $\Gamma-$X axes of
the Brillouin zone at $|{\bf k}_\mu|\sim 0.85(2\pi/{\rm a})$ from the
$\Gamma$ point]:
\begin{equation}
\psi_{{\bf R}_0} ({\bf r}) = \frac{1}{\sqrt{6}}\sum_{\mu = 1}^6  F_{\mu}({\bf
r}-
{\bf R}_0) u_\mu({\bf r}) e^{i {\bf k}_{\mu}\cdot ({\bf r}-{\bf R}_0)}\,.
\label{eq:sim}
\end{equation}
The phases of the plane-wave part of all band edge Bloch states are naturally
chosen to be pinned at ${\bf R}_0$: In this way the charge density at the
donor site [where the donor perturbation potential energy Eq.~(\ref{eq:coul})
is the smallest] is maximum, thus minimizing the energy for $\psi_{{\bf
R}_0}({\bf r})$.

In Eq.~(\ref{eq:sim}), $F_{\mu}({\bf r}-{\bf R}_0)$ are envelope functions 
centered at ${\bf R}_0$, for which we adopt the anisotropic Kohn-Luttinger
form, e.g., for $\mu = z$, 
$F_{z}({\bf r}) = \exp\{-[(x^2+y^2)/a^2 + z^2/b^2]^{1/2}\}/\sqrt{\pi a^2 b}$.
The effective Bohr radii $a$ and $b$ are variational parameters chosen to
minimize $E_{SV} = \langle\psi_{{\bf R}_0}| {\cal H}_{SV} |\psi_{{\bf
R}_0}\rangle$, leading to $a=25$ \AA, $b=14$ \AA~ and $E_{SV} \sim -30$ meV
when recently measured effective mass values are used in the
minimization \cite{KHD1}.  The periodic part of each Bloch function is pinned
to the lattice, independent of the donor site.  

The ${\cal H}_{SV}$ ground state is six-fold degenerate due to the six-fold
valley degeneracy of Si conduction band.  This degeneracy is lifted by the
valley-orbit interactions \cite{Pantelides}, which are included here in
${\cal H}_{VO}$, leading to the nondegenerate ($A_1$-symmetry) ground state
in (\ref{eq:sim}).

The HL approximation is a reliable scheme for the well-separated donor pair
problem (interdonor distance much larger than the donor Bohr radii)
\cite{slater}. Within HL, the lowest energy singlet and triplet wavefunctions
for two electrons bound to a donor pair at sites $\mathbf{R}_A$ and
$\mathbf{R}_B$, are written as properly symmetrized combinations of
$\psi_{\mathbf{R}_A}$ and $\psi_{\mathbf{R}_B}$ [as defined in
Eq.(\ref{eq:sim})]
\begin{equation}
\Psi^s_t({\mathbf r}_1,{\mathbf r}_2) = \frac{1}{\sqrt{2(1\pm S^2)}}
\left[ \psi_{{\bf R}_A}({\bf r}_1) \psi_{{\bf R}_B}({\bf r}_2) \pm 
\psi_{{\bf R}_B}({\bf r}_1) \psi_{{\bf R}_A}({\bf r}_2) \right],
\label{eq:hl}
\end{equation}
where $S$ is the overlap integral and the upper (lower) sign corresponds to
the singlet (triplet) state.  The energy expectation values for these states, 
$E^s_t = \langle\Psi^s_t|{\cal H}|\Psi^s_t\rangle$, give the exchange
splitting through their difference, $J=E_t-E_s$.  We have previously derived
the expression for the donor electron exchange splitting \cite{KHD2,KCHD},
which we reproduce here:
\begin{eqnarray} 
J({\bf R}) = \frac{1}{36}\sum_{\mu, \nu} {\cal J}_{\mu \nu}
({\bf R}) \cos ({\bf k}_{\mu}-{\bf k}_{\nu})\cdot {\bf R}\,,
\label{eq:exch}
\end{eqnarray}
where $\mathbf{R} = \mathbf{R}_A - \mathbf{R}_B$ is the interdonor position
vector and ${\cal J}_{\mu \nu} ({\bf R})$ are kernels determined by the
envelopes and are slowly varying \cite{KHD1,KHD2}.  Note that
Eq.~(\ref{eq:exch}) does not involve any oscillatory contribution from 
$u_{\mu}({\bf r})$, the periodic part of the Bloch functions
\cite{wellard03,KCHD}.  The physical reason for that is clear from
(\ref{eq:sim}): While the plane-wave phases of the Bloch functions are pinned
to the donor sites, leading to the cosine factors in (\ref{eq:exch}), 
the periodic functions $u_\mu$ are pinned to the lattice, regardless of the
donor location. 

\begin{figure}
\includegraphics[width=4.1in]
{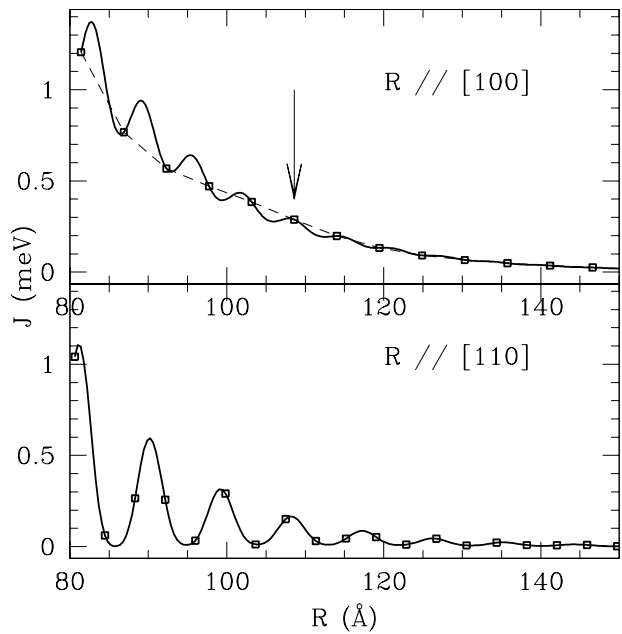}
\protect\caption[Two frames with oscillatory exchange]
{\sloppy{
Exchange coupling between two phosphorus donors in Si along the indicated
directions in the diamond structure. Values appropriate for impurities at
substitutional sites are given by the squares.
The dashed line in the $R\|[100]$ frame is a guide to the eye, indicating that
the oscillatory 
behavior may be ignored for donors positioned exactly along this axis.
}}
\label{fig:oscillate}
\end{figure}

The exchange energy calculated from Eq.~(\ref{eq:exch}) for a pair of donors 
as a function of their relative position along the [100] and [110] crystal
axis is given in Fig.~\ref{fig:oscillate}.
This figure vividly illustrates both the anisotropic and the oscillatory
behavior of $J({\bf R})$, which is well established from previous studies
\cite{Andres,KHD1,KHD2,wellard03}.  It is interesting to note that for
substitutional donors with interdonor position vectors exactly aligned with
the [100] crystal axis, the oscillatory behavior may be ignored in practice,
as indicated by the dashed line in the figure.  This behavior is
qualitatively similar to the exchange versus donor separation dependence
assumed in Kane's proposal \cite{Kane}, where the Herring and Flicker
expression \cite{herring64}, originally derived for H atoms,  was adapted for
donors in Si.  Therefore one might expect that reliable exchange gate
operations would be possible if all donor pairs are exactly aligned along the
[100] crystal axis.

\subsection{Nanofabrication aspects}

\begin{figure}
\includegraphics[width=4.1in]
{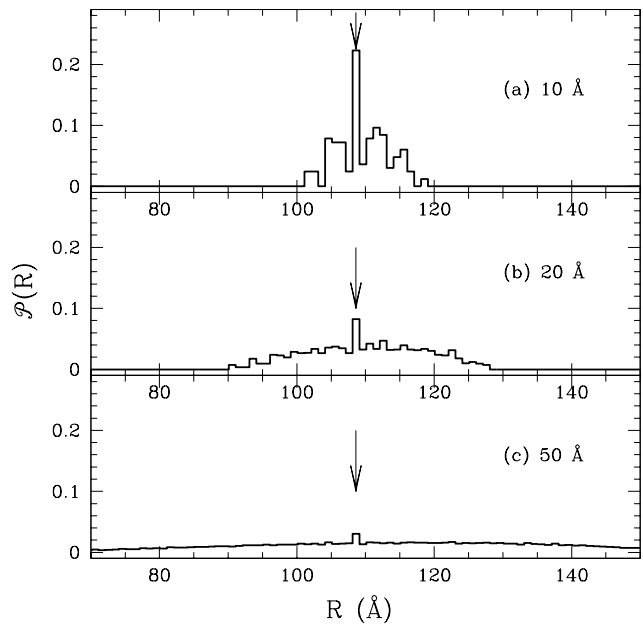}
\protect\caption[3 frames with R distributions]
{\sloppy{
Interdonor distance distributions for a {\it target} relative position of 20 
lattice parameters along [100] (see arrows). The first donor is fixed and the
second one ``visits'' all of the Si substitutional lattice sites within a
sphere centered at the {\it target} position, with uncertainty radii (a)10
\AA, (b)20 \AA~and (c)50 \AA. 
}}
\label{fig:histo}
\end{figure}

Aiming at the fabrication of a P donor array accurately positioned along the
[100] axis, and taking into account the current state of the arts degree of
control in substitutional P positioning in Si of a few nm
\cite{Obrien,encapsulation,implant,schenkel03}, we investigate the
consequences of such interdonor positioning uncertainties ($\sim$ a few nm)
in the values of the corresponding pairwise exchange coupling.  We define the
{\it target} interdonor position ${\bf R}_t$ along [100], with an arbitrarily
chosen length of 20 lattice constants $(\sim 108.6$ \AA) indicated by the
arrow in Fig.~\ref{fig:oscillate}.  The distributions for the interdonor
distances $R = |{\bf R}_A-{\bf R}_B|$ when ${\bf R}_A$ is fixed and ${\bf
R}_B$ ``visits'' all of the diamond lattice sites within a sphere centered at
the {\it target} position are given in Fig.~\ref{fig:histo}. Different frames
give results for different uncertainty radii, and, as expected, increasing
the uncertainty radius results in a broader distribution around the {\it
target} distance. Note that the geometry of the lattice implies that the
distribution is always centered and peaked around $R_t$, as indicated by the
arrows.  The additional peaks in the distribution reveal the discrete nature
of the Si lattice.
\begin{figure}
\includegraphics[width=4in]
{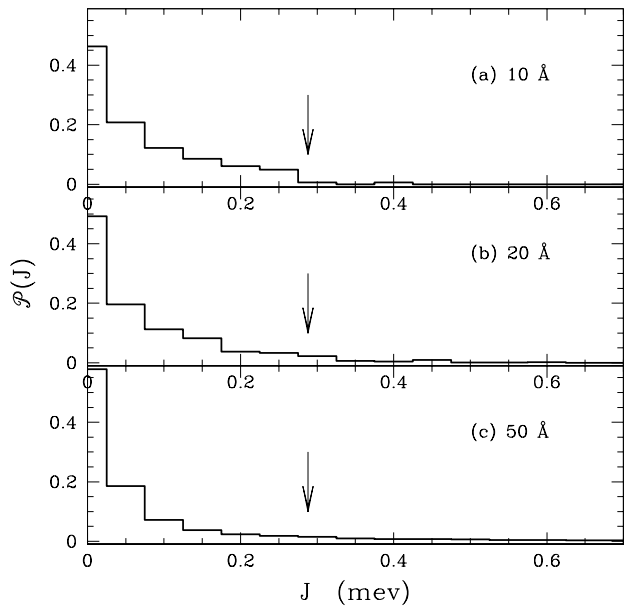}
\protect\caption[3 frames with exchange distributions]
{\sloppy{
Exchange distributions for the same relative position {\it ensembles} in
Fig.~\ref{fig:histo}.  The arrow indicates the {\it target} situation.
Contrary to the distance distributions, the exchange distributions are not
centered or peaked around the target value.
}}
\label{fig:histoj}
\end{figure}

The respective distributions of exchange coupling between the same donor pairs 
in each {\it ensemble} is presented in Fig.~\ref{fig:histoj}, where the arrows
give the exchange value at the {\it target} relative position: $J({\bf R}_t)
\sim 0.29$ meV.  The results here are qualitatively different from the
distance distributions in Fig.~\ref{fig:histo}, since they are neither
centered nor peaked at the {\it target} exchange value. Even for the smallest
uncertainty radius of 1 nm in (a), the exchange distribution is peaked around
$J\sim 0$, bearing no semblance to the inter-donor distance distributions.
Increasing the uncertainty radius leads to a wider range of exchange values, 
with a more pronounced peak around the lowest $J$ values.

From the perspective of current QC fabrication efforts, $\sim 1$ nm accuracy
in single P atom positioning has been recently demonstrated
\cite{encapsulation}, representing a major step towards the goal of obtaining
a regular donor array embedded in single crystal Si.  Distances and exchange
coupling distributions consistent with such accuracy are presented in
Figs.~\ref{fig:histo}(a) and \ref{fig:histoj}(a) respectively.  The present
calculations indicate that even such small deviations ($\sim$ 1 nm) in the
relative position of donor pairs with respect to perfectly aligned
substitutional sites along [100] lead to order-of-magnitude changes in the
exchange coupling, favoring $J\sim 0$ values.  Severe limitations in
controlling $J$ would come from ``hops'' into different substitutional
lattice sites.  Therefore, precisely controlling of exchange gates in Si
remains an open (and severe) challenge.

\subsection{Strained Si}

\begin{figure}
\includegraphics[width=4in]
{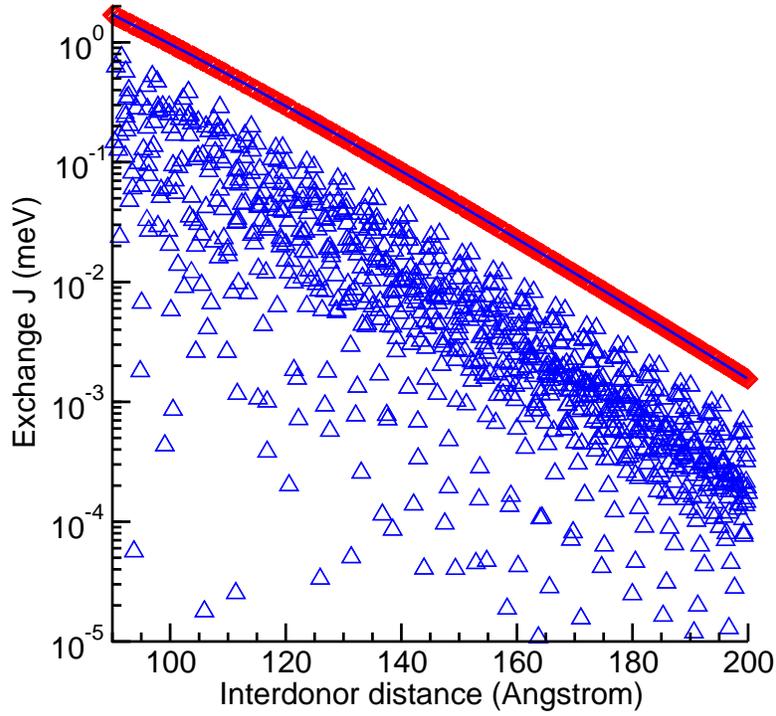}
\protect\caption[Exchange in-plane]
{\sloppy{
Calculated exchange coupling for in-plane displacements of the donors in the
$x$-$y$ plane for relaxed Si ($\chi=0$, with diamond symbols) and strained Si
(uniaxial strain along the $z$ direction with $\chi = -20$, with square
symbols forming a straight line).  The relative positions considered for the
donor pairs consist of one donor at all possible lattice sites between two
concentric circles of radii 90 \AA~ and 180 \AA~ with the other donor
positioned at the center of the circles.  The data points correspond to the
exchange calculated at all relative positions considered.  The solid line is
$J({\bf R})$ for ${\bf R}$ along the [100] direction for $\chi = -20$.
}}
\label{fig:Exch-plane}
\end{figure}

\begin{figure}
\includegraphics[width=4in]
{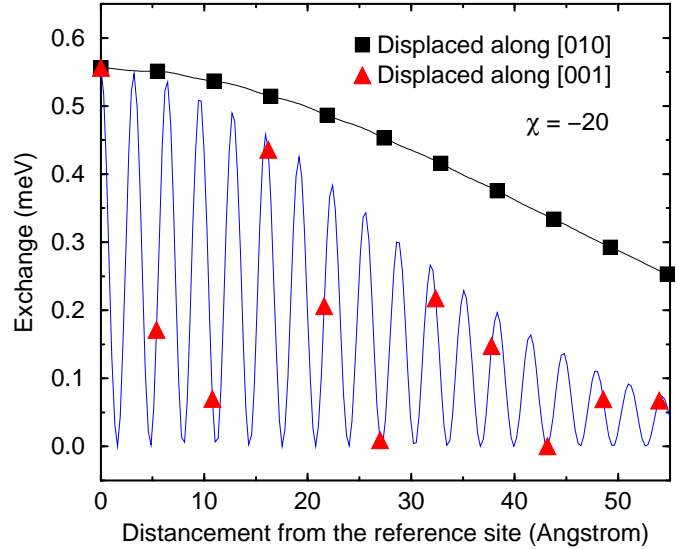}
\protect\caption[Exchange in-plane]
{\sloppy{
Donor electron exchange splitting in Si uniaxially strained along
[001] direction ($\chi=-20$, corresponding to the strain in a Si quantum well
grown in between relaxed Si$_{0.8}$Ge$_{0.2}$ barriers).  The two donors are
approximately aligned along the [100] direction, with one of them displaced
along the [010] direction (original figure appeared in
Ref.~\onlinecite{KHD2}).
}}
\label{fig:Exch-[010]}
\end{figure}

Uniaxial strain can be used to break the Si lattice symmetry and partially
lift the degeneracy between the valleys, so that as few as two valleys make
up the bottom of the conduction band.  Then the sum over $\mu$ and $\nu$ in
Eq.~(\ref{eq:exch}) is much simplified, but the sinusoidal factor will still
remain, so that care still has to be taken in controlling the donor exchange
\cite{KHD2}.  For example, if a uniaxial strain is applied along $z$
direction, variation of relative donor position in the $x$-$y$ plane does not
lead to oscillation in the inter-donor exchange, as shown in
Fig.~\ref{fig:Exch-plane}.  Strain is quantified here by a dimensionless
parameter, $\chi$, defined in Ref.~\cite{KHD2}.  On the other hand, atomic
scale donor movement in a direction parallel to the strain direction $z$ can
still cause order-of-magnitude change in the donor exchange coupling, as is
illustrated in Fig.~\ref{fig:Exch-[010]}.  Therefore the exchange oscillation
problem (leading to essentially random variations in the exchange coupling
arising from dopant positioning within the Si unit cell) remains even in
uniaxially strained Si since the system still has a two-valley degeneracy. 
These theoretical considerations \cite{KHD2} have recently been verified in a
more sophisticated approximation which took into account higher-order Coulomb
interaction corrections \cite{wellard03}.

Given the random variations in the exchange energy as a function of
unintentional (and unavoidable at present) small ($\sim$ 1 nm) variations in
the positioning of P dopants in Si unit cell, as shown in
Figs.~\ref{fig:oscillate}-\ref{fig:Exch-[010]}, a careful rethinking of the
exchange gate fabrication and control for Si QC architecture may be required. 
This is particularly true in view of the fundamental origin of the exchange
oscillations, which arises from the quantum interference between the
degenerate valleys in the silicon conduction band and therefore cannot be
eliminated by simple fabrication tools (unless, of course, dopant positioning
with sub-nanometer precision on the Si lattice somehow becomes available). 
One possibility is to eliminate the use of exchange gates and instead rely
completely on hyperfine coupling and electron shuttling \cite{Skinner}.
Another possibility could be to go away completely from shallow donor bound
states in Si and instead use SiGe quantum dot structures to confine single
electrons, making the Si QC essentially identical to the proposed GaAs
quantum dot QC \cite{Friesen}.  Even in this case, however, the valley
degeneracy problem may remain unless the quantum dot confinement potential is
extremely weak (which would cause other problems such as the low lying
orbital excited states potentially jeopardizing the two-level dynamics of the
spin qubit).  In any case, our results of the strong (essentially random)
qubit-to-qubit variations in the exchange energy in the P-doped
shallow-donor-based Si QC architecture compromises one of the main perceived
advantages of Si QC (over, for example, the GaAs quantum dot QC), namely, the
identical nature of each spin qubit (arising from all P shallow donor states
in Si being identical by definition) does not seem to translate to similar
exchange gate characteristics--in fact, the exchange gate turns out to be
essentially random in nature (with wide variations in its strength peaking at
zero). 

Comparing with Si, a question naturally arises about the GaAs quantum dot QC
where the exchange would also exhibit random qubit-to-qubit variations
arising from the (essentially trivial) random variations ($\pm$10\%-20\%) in
quantum dot confinement sizes and inter-dot separations since it is
impossible to fabricate identical quantum dots and/or to place them perfectly
in an architecture.  The quantum dot exchange variation problem (arising in
the leading order from the $e^{-R/a}$ sensitivity of the exchange energy of
the inter-dot separation $R$) is, however, less severe because the ``random''
exchange in this case will be peaked around a finite value in contrast to Si
QC, where the most probable value of the inter-qubit exchange coupling is
zero.  We note, however, that in both cases one may have to carry out
extensive characterization of the individual qubit-to-qubit coupling strength
in order to operate the exchange gate, and such qubit characterization itself
may turn out to be a hard control problem.  We have recently proposed an
experimental technique of spatially resolved micro-Raman spectroscopy as 
possible diagnostic tool to characterize local values of exchange coupling
between individual spin qubits (i.e. the singlet-triplet energy splitting for
individual pairs of induced shallow electron donor states) \cite{KHDD}.

\section{Summary}

In summary, we have briefly reviewed two important issues related to
donor-spin based quantum computing in silicon: quantum coherence of electron
spins and exchange interaction among donor electrons.  Our results show that
the spin qubits based on donors in silicon have remarkable potential for very
long quantum coherence times through isotopic purification.  On the other
hand, they also pose immense challenges in terms of precise nanostructure
fabrications because of the degenerate nature of the silicon conduction band,
which leads to essentially a random exchange gate coupling.  Further studies
of fabrication and innovative alternative approaches are imperative in order
to fully realize the potential of this donor-based QC architecture.  We have
discussed an interesting dichotomy between donor-based Si and quantum
dot-based GaAs QC architectures.  In particular, isotopic purification, which
can eliminate essentially all free nuclear spins (by eliminating $^{29}$Si
nuclei) in silicon (but not in GaAs), gives Si an enormous advantage of very
long ($T_2 \gtrsim 10-100$ ms) electron spin decoherence time compared with
the corresponding GaAs quantum dot situation ($T_2 \sim 10-100$ $\mu$s, three
orders of magnitude shorter than in Si), making Si an ideal QC candidate
material.  However, the valley degeneracy of Si conduction band leads to
demands for extremely precise dopant positioning (so that the inter-qubit
exchange coupling is finite for most qubits) which will be difficult to
achieve.  It may be worthwhile in this context to consider alternative
two-qubit gates, such as the dipolar gate \cite{DD_dipolar} or electron
shuttling \cite{Skinner}, for silicon quantum computation.

\begin{acknowledgments}
This work was supported by ARDA and LPS-NSA, as well as
by CNPq, Instituto do Mil\^enio de
Nanoci\^encias  and FAPERJ in Brazil, by ARO-ARDA at the
University at Buffalo and the University of Maryland, and by DARPA SpinS
program at the University of California at Berkeley. 
\end{acknowledgments}

\end{document}